# Spectrally resolving the phase and amplitude of coherent phonons in the charge density wave state of 1$T$-TaSe$_2$

Charles J. Sayers[1], Stefano Dal Conte[1], Daniel Wolverson[2], Christoph Gadermaier[1,3], Giulio Cerullo[1], Ettore Carpene[4] and Enrico Da Como[2]*

1 Dipartimento di Fisica, Politecnico di Milano, 20133 Milano, Italy

2 Department of Physics, University of Bath, BA2 7AY Bath, United Kingdom

3 Center for Nano Science and Technology @PoliMi, Istituto Italiano di Tecnologia, 20133 Milano, Italy

4 IFN-CNR, Dipartimento di Fisica, Politecnico di Milano, 20133 Milano, Italy

**Abstract**

The excitation and detection of coherent phonons has given unique insights into condensed matter, in particular for materials with strong electron-phonon coupling. We report a study of coherent phonons in the layered charge density wave (CDW) compound 1$T$-TaSe$_2$ performed using transient broadband reflectivity spectroscopy, in the photon energy range 1.75-2.65 eV. Several intense and long lasting (> 20 ps) oscillations, arising from the CDW superlattice reconstruction, are observed allowing for detailed analysis of the spectral dependence of their amplitude and phase. We find that for energies above 2.4 eV, where transitions involve Ta $d$-bands, the CDW amplitude mode at 2.19 THz dominates the coherent response. At lower energies, instead, beating arises between additional frequencies, with a particularly intense mode at 2.95 THz. Interestingly, our spectral analysis reveals a $\pi$ phase shift at 2.4 eV. Results are discussed considering the selective coupling of specific modes to energy bands involved in the optical transitions seen in steady-state reflectivity. The work demonstrates how coherent phonon spectroscopy can distinguish and resolve optical states strongly coupled to the CDW order and provide additional information normally hidden in conventional steady-state experiments.

*email: edc25@bath.ac.uk

## 1. Introduction

The photoexcitation of coherent electronic and phonon modes by ultrashort laser pulses (< 100 fs) is a very active and fascinating field of condensed matter research[1]. Over the years, fundamental discoveries such as the measurement of coherent vibrational wave packets in retinal molecules responsible for human vision[2], the detection of coherent Bloch oscillations in coupled semiconductor quantum wells[3] and the demonstration of THz radiation emission[4], to name a few, have shown the versatility of the technique and the potential for opening new areas, with the ultimate aim of controlling vibrational and electronic degrees of freedom[5]. Recently, coherent phonon spectroscopy has been extensively applied to correlated electron systems and quantum materials[6]. These are solids showing superconductivity, Mott insulator transitions, charge density wave (CDW) instabilities and, in general, strong coupling between electronic and spin degrees of freedom with the lattice dynamics. The widespread suitability of femtosecond spectroscopy is demonstrated in several examples of excitation and detection of coherent phonons in CDW materials[7].

In a CDW, strong coupling of electronic bands close to the Fermi level with specific phonon modes can be responsible for a structural phase transition, usually occurring upon cooling below some critical temperature. The phase transition results in the opening of an electronic gap in materials that normally would be metals or semimetals and the formation of a superlattice indicated by the softening of specific phonons[8]. The superlattice structure folds the electronic and phonon dispersion into a new reduced Brillouin zone (BZ), resulting in a significant renormalization of the phonon spectrum[8a].

We focus on the CDW material 1*T*-TaSe$_2$ which belongs to the family of quasi-2D layered transition metal dichalcogenides (TMDs)[9] and is characterized by a relatively high transition temperature compared to the closely related compound 1*T*-TaS$_2$[10] and other metallic TMDs known to exhibit CDWs[11]. Below Tc = 470 K, a first order transition occurs from an incommensurate to a commensurate CDW with long range order[10]. Thus, at room temperature and below, the material is in the CDW superlattice structure characterized by Ta atoms displaced from their normal lattice sites and clustering into 13 atom stars as indicated in Figure 1(a). In terms of electronic structure, the CDW superlattice results in band folding and the opening of gaps, which are localized on specific portions of the Fermi surface. This leaves states at the Fermi level, which contribute to conduction and can explain the overall metallic behaviour seen in resistivity[8a, 12]. Together with the CDW, other electronic correlations arising from Mott physics have been reported in experiments sensitive to the surface electronic structure or in monolayer samples[7d, 13]. Figure 1 (b) is an approximate sketch of the local density of states (DOS) drawn considering the recent data obtained by Crommie and coworkers on trilayer 1*T*-TaSe$_2$[13b]. The vertical arrows give an idea of the energy bands possibly involved in optical transitions within the photon energy interval used in our experiments. The photon energies used in our experiments are well above the energy separation between the empty and occupied electronic bands close to the Fermi level, here indicated as VB1 and CB1. The density of states at VB2 and lower energies has a strong contribution from the chalcogen 2*p* band[7d] similar to 1*T*-TaS$_2$[14], while VB1 and CB1 are mainly due to Ta *d* orbitals.

Recent experiments on 1*T*-TaSe$_2$ by femtosecond time- and angle-resolved photoemission spectroscopy (TR-ARPES)[7d, 12, 15], have highlighted how the binding energy of VB1 is modulated by a phonon corresponding to the breathing mode of the star cluster. This is revealed by long lasting oscillations seen in the TR-ARPES signal at ~2.2 THz. However, the CDW superlattice reconstruction (√13 x √13) not only results in the breathing mode, but several acoustic and optical phonon branches of the normal lattice becoming optically active because of zone folding, as seen in time resolved reflectivity (TRR) and spontaneous Raman experiments[7d, 16]. Thus, it remains unclear why only some of the phonon modes of the superlattice are coupled with the CDW order probed directly in TR-ARPES. In addition, the nature of the coherent excitation process, i.e., displacive versus impulsive, remains undetermined. Phase information, crucial to understanding the phonon excitation process, relies on quality of time resolved data and exact determination of the temporal overlap between pump and probe pulses. TRR is highly suitable for this purpose because of the availability of very short pulses (<20 fs) and the methods required to monitor and characterize pulses in the visible and NIR range, in contrast to deep UV used for TR-ARPES.

Here, we report high quality TRR data on 1$T$-TaSe$_2$ single crystals at 77 K. Benefitting from a broadband visible probe pulse, spanning 1.75 - 2.65 eV, we are able to track the dynamics of different optical interband transitions, study the selective phonon modulation of transient reflectivity signal, and retrieve important information about the phase of coherent phonon oscillations. In contrast to TR-ARPES, we are able to detect several coherent phonons linked to the CDW superlattice reconstruction and investigate their spectral dependence. Interband transitions above 2.4 eV are coupled mainly, to the CDW amplitude breathing mode, while below this photon energy beating of several comparably intense modes is observed. Phase analysis can distinguish modes excited via a displacive or impulsive process, with the former preferentially coupled to optical interband transitions. We observe a π phase shift of the displacive CDW amplitude modes in correspondence to an optical transition at 2.4 eV. Our results show how broadband coherent phonon TRR spectroscopy can combine high energy and time resolution to provide unique insights into the electron and lattice dynamics of CDW materials. In addition, by comparing TRR with the available TR-ARPES data we discuss the differences in the detection of coherent phonons in the two techniques and how these can be related to the electronic bands involved in the CDW formation.

## 2. Results

The ultrafast optical response of 1$T$-TaSe$_2$ is characterized by intense long lasting coherent oscillations. Figure 1(c) reports the temporal evolution of the broadband differential reflectivity signal ($\Delta$R/R) of the probe pulse following the excitation of a 1$T$-TaSe$_2$ single crystal held at 77 K. The two-dimensional data map is dominated by oscillations which remarkably last longer than our experimental time window of 25 ps. Time traces at specific probe photon energies are reported in Fig.1(d) and show how at 2.6 eV the dynamics is dominated by a single strong oscillatory mode. This is in contrast to traces at energies of 2.35 eV and 1.80 eV, where beating of modes is detected.

Figure 1(e) shows $\Delta$R/R spectra extracted at specific time delays. At early positive delays of 0.32 ps and of 2 ps spectra show a negative band peaking around 2.6 eV, which develops into a positive $\Delta$R/R signal as photon energy decreases. The spectra also feature maxima at positive signal in the region below 2.3 eV. It is important to mention that the spectral position at which the $\Delta$R/R changes sign oscillates in time. This can be seen by the two short delay spectra and also the border between red and blue regions in the two-dimensional map of Fig.1(c). A closer inspection of the spectral dynamics for all the delays reveals that the positive peak in $\Delta$R/R at ~2.1 eV oscillates mainly in intensity, while the negative peak at 2.6 eV oscillates also in energy with a shift of as much as ± 0.2 eV. Such behaviour already suggests that the two spectral regions above and below 2.4 eV have a different coupling with the oscillatory modes and likely a different nature. This separation is more evident after subtracting the fast incoherent decay from the data (details in the Methods) in order to isolate the oscillatory component as shown in Figure 2(a). Here, the presence of a dominant mode at energies above 2.4 eV and of a beating below become clearer (Fig.2 (b)). The nodes in Figure 2(a) arising from beating appear as white vertical stripes at regular intervals, which do not extend to the region of probe photon energy above 2.4 eV.

We have performed a detailed analysis of the oscillations by a Fourier transform of the time traces at the different probe photon energies as reported in Figure 3(a). There are at least four distinct oscillation frequencies clearly visible which contribute to the dynamics. From Figure 3(b) their frequencies are 1.61 THz, 1.82 THz, 2.19 THz and 2.95 THz and correspond to Raman active phonon modes from the CDW superstructure[16]. The most prominent oscillation at 2.19 THz is a totally symmetric $A_g$ breathing mode of the star cluster, also known as the amplitude mode of the CDW superlattice. The 2.95 THz feature has been also assigned to an $A_g$ mode of the CDW, while the remaining two at lower frequency are close in energy to two almost degenerate CDW phonons with $A_g$ and $E_g$ symmetry[17]. The $A_g$ mode is the more likely assignment given that the TRR signal is dominated by such modes because of symmetry reasons[18]. A scheme illustrating the vibrations of the 13 atoms in the star cluster for the two most intense modes is shown in the inset. These were obtained from density functional theory (DFT) calculations on a 1$T$-TaSe$_2$ monolayer in the CDW phase. The Supplementary Information (SI) outlines the details of the calculations together with the assignment of phonon modes. Differences between the frequency of calculated and measured phonons are due to the harmonic approximation used in DFT and possibly because of the monolayer input structure compared to the bulk, as measured in the experiments.

The Fourier map in Fig.3(a) clearly shows that for energies greater than 2.4 eV, the 2.19 THz amplitude mode is dominant with a small contribution from the 1.82 THz phonon. The intensity of the amplitude mode is negligible for a small interval close to 2.4 eV and remerges with a relative maximum at ~2.2 eV. The phonon at 2.95 THz, in contrast, is almost absent above 2.4 eV and becomes dominant with a peak at probe energy 2.17 eV. The amplitude of the four modes as a function of probe photon energy, corresponding to vertical cuts on the map, can be seen in Fig.3(d) providing an immediate identification of 2.4 eV as the energy where a change in the optical properties occurs.

One of the unique advantages of coherent phonon spectroscopy compared to the spontaneous Raman scattering is the possibility to retrieve phase information. We report in Fig.3(c) how the modes' phases change as a function of probe photon energy. Interestingly, the amplitude mode shows an abrupt phase shift of $\pi$ occurring in proximity of 2.4 eV; this is a clear signature of an optical transition involving at least one electronic band with binding energy that changes in time as we explain in the following. The high frequency mode at 2.95 THz shows also a $\pi$ phase shift, albeit less abrupt and occurring over a wider energy range. The phase behaviour of the two modes at lower frequency is different, since the phase of the 1.61 THz is very close to $\pi$, while for the 1.82 THz mode phase is almost constant across the spectrum at $\pi/2$, implying an impulsive excitation for the latter.

To gain insight into the relationship between coherent phonons and optical transitions we have compared the time-resolved data with the steady state optical reflectivity of 1$T$-TaSe$_2$. The reflectivity spectrum of a single crystal is reported in Figure 4(a). The spectrum displays two clear minima centred at 3.2 eV and 2.7 eV. In addition, the increasing reflectivity towards low photon energy features a shoulder at 2.1 eV. To the best of our knowledge, this is the first time that the reflectivity in the visible spectral range for this material is reported. The overall shape reproduces features theoretically predicted by Reshak and Auluck[19] and is similar to the one of 1$T$-TaS$_2$ measured by Beal et al.[20] despite a small rigid energy shift. The reflectivity increase observed when going towards the infrared could be due to an interband transition centred just outside our spectral window superimposed to the long tail of a Drude free carrier response, given that the compound shows metallic behaviour down to 4 K.

The reflectivity data on single crystals of the isostructural CDW compound 1$T$-TaS$_2$, measured from 25 meV to 14 eV, indicate that the increased reflectance towards low energy is indeed due to free carriers[20], while the optical transitions in the region above 2.4 eV are assigned to interband transitions involving bands with Ta $d$ character. Since Ta bands are close to the Fermi level, it is very likely that optical transitions in the spectral range of our probe involve promotion of electrons from the VB1 (initial state) or to the CB1 (final state), as shown in the energy scheme of Fig. 1(b).

The first derivative of the steady state reflectivity spectrum, -δR/δE, in Fig. 4(a) highlights the spectral features with increased clarity. Interestingly the greatest change in reflectivity is observed at 2.3 eV, which is very close to the energy where the most abrupt changes in phonon dynamics are observed. A comparison with the ΔR/R spectra at different delays in Fig. 4(b) shows the similarity of spectral features with the first derivative of the steady state reflectivity spectrum.

## 3. Discussion.

One of the goals of our experiments is to understand the coupling of coherently excited phonons with the optical transitions observed in 1$T$-TaSe$_2$. We stress that our pump-probe experiments are in the weakly perturbative regime, since a laser fluence of 10 µJ/cm$^2$ corresponds to an excitation density well below the threshold for photoinduced melting of the CDW order and phase transition to an incommensurate CDW or metastable states reported in ref. [15].

We interpret the interesting behaviour seen in the phonon oscillations benefitting from the DFT calculations performed on a monolayer 1$T$-TaSe$_2$ in the CDW structure, i.e. with star clusters arranged in a ($\sqrt{13}$ x $\sqrt{13}$) hexagonal superlattice as shown in Fig.1(a). The vectors representing the atomic displacement in the star cluster for different phonon modes are shown in the insets of Figure 3. A full list of phonon modes and their symmetries is provided in the SI. The 2.19 THz mode corresponds to the expansion and contraction of the 12 atoms surrounding the central Ta atom of the star, thus modulating the CDW superlattice structure. In addition, the mode assigned to the 2.95 THz oscillation show radial displacements, but predominantly involving the six atoms at the tips of the star and only small amplitude motions for the six nearest neighbours to the centre. The assignment of the oscillations at 1.61 THz and 1.82 THz is non-trivial based on their frequency but is close to the A$_g$ modes seen in Raman[17]. We propose the assignment given in the SI considering that calculated low frequency A$_g$ modes start with 1.4 THz and then 1.82 THz.

The CDW amplitude mode at 2.19 THz is directly linked to the lattice reconstruction, and thus coupled to the physics of Ta-$d$ bands VB1 and CB1 as suggested for the similar compound 1$T$-TaS$_2$[21]. Our data showing a π phase shift for this mode at 2.4 eV, can be interpreted considering an optical transition occurring at this energy and involving either VB1 or CB1. Oscillations of VB1 is consistent with our TR-ARPES data, where we have observed a 2.19 THz modulation of the band edge energy[7d].

The oscillations of either bands lead to a ΔR/R signal above 2.4 eV being one half-cycle (π shift) out of phase from the signal just below, as illustrated in Fig.4(c). Similar arguments can explain the phase behaviour of the 2.95 THz oscillation, which at t = 0 is out of phase by π with the amplitude mode and shows a more progressive change of π as a function of photon energy. Its amplitude is enhanced for optically coupled states at energies below 2.4 eV (cf. Fig.3(c)), thus we speculate it may be linked to a different transition, possibly involving the CB1. The oscillations at 1.82 THz have a phase of |π/2| indicating an impulsive excitation. Their intensity should be proportional to the Raman tensor and its dependence on photon energy[22]. The negligible amplitude of the 1.61 THz mode below 2.4 eV is indicative of preferential coupling with optical transitions similar to those of the amplitude mode.

In general, coherent phonon oscillations seen in TRR of absorbing materials depend on the derivative of the complex susceptibility with respect to the atomic coordinates[22]. Thus, the coherent phonon oscillations probed in our experiments here are not solely linked to the dynamics of electronic bands probed by TR-ARPES[7d, 15], but rather a convolution of energy shifts in bands and changes in electronic susceptibility. TR-ARPES studies on 1T-TaSe$_2$ have so far focussed on the dynamics of VB1 close to the Γ point and little is known about CB1 and other empty bands which are suspected to also contribute to the signal detected in TRR experiments.

In TR-ARPES the signal from VB1 is dominated by the amplitude mode at 2.19 THz[7d], while TRR shows at least four different modes. Our analysis of phase in Fig.3c shows that together with the amplitude mode also the 1.61 THz and 2.95 THz have a displacive nature and are thus excited in connection to an interband transition. Of these two, only the 2.95 THz shows a sluggish $\pi$ phase shift occurring over a broad energy range between 2.2 eV and 2.6 eV.

We can interpret the TRR results considering that the amplitude mode at 2.19 THz directly modulates the binding energy of VB1 and that this band is likely the initial state of an optical transition at 2.4 eV. According to TR-ARPES the 1.61 THz and 2.95 THz do not modulate the binding energy of VB1, but they may modulate other empty bands which are the final states of other optical transitions, for example for optical transitions at energies below 2.4 eV. Conduction bands (CBs) are not easily seen in TR-ARPES because of their fast depopulation by relaxing electrons. Often his does not allow to follow the influence of coherent phonons on the binding energy of CBs on time scales sufficiently long to measure oscillatory signals. Lastly, we stress again that coherent phonon oscillations in TRR are not necessarily linked to changes in the binding energy of electronic bands but arise from the modulation of the electronic susceptibility by specific phonon modes. Therefore, the oscillations at 1.61 THz and 2.95 THz may be hard to detect by a technique sensitive mainly to binding energy shifts and electron density such as TR-ARPES. Figure 4(c) is a schematic illustration of how the CDW amplitude mode can modulate the binding energy of an initial state at energy $E_i$ and final state at energy, Ef, involved in an optical transition. Optical signals at photon energies above (blue arrow) and below (red arrow) $|E_f - E_i|$ will oscillate with a difference in phase of half-cycle, $\pi$. As shown in Figure 4(b) changes in spectral position of the $\Delta R/R$ minimum at ~2.6 eV can be up to 0.2 eV. Less dramatic shifts are observed for the peak at 2.1 eV, confirming the different nature of optical transitions in this region. An energy shift up to 0.2 eV is inconsistent with a modulation of VB1 binding energy of only ~50 meV seen in TR-ARPES for a higher pump fluence[7d]. This observation rules out that oscillations in VB1 binding energy are solely responsible to what is seen in TRR and coherent oscillations in final states in the conduction band and other phenomena shifting spectral weight must be at play. A more complete description requires models beyond single particle energy bands and considering electronic correlations.

In summary, we have performed a detailed study of coherent phonons in 1$T$-TaSe$_2$ by femtosecond TRR spectroscopy. The results show several coherently coupled phonons, some of them giving a strong periodic modulation to the optical signal in the visible range. The spectrally resolved phonon dynamics including both amplitude and phase allows for a better identification of optical transitions compared to steady state optical reflectivity, where for CDW materials based on TMDs the signal of interband transitions is often merged with a broad background due to Drude-type intraband optical phenomena[23]. The results are also useful towards the interpretation of recent TR-ARPES experiments claiming a shift in the phase between oscillations in band binding energy and electronic temperature[24] as well as calling for a better understanding of the occupied and empty bands close to the Fermi level which have been linked to Mott physics and electronic correlations in this material[13b].

**Methods.**

Single crystals of 1$T$-TaSe$_2$ were grown using the Iodine vapour transport method in vacuum sealed quartz ampules. Rapid quenching from 920 °C to room temperature in a water/ice bath ensured the stabilization of the octahedral 1$T$ polytype. Further sample details and basic structural characterization of the single crystals are reported in a previous publication[7d]. A freshly cleaved single crystal was mounted inside an optical flow cryostat and held at 77 K. Transient reflectivity experiments were performed using a setup based on a Ti:sapphire amplified laser (Coherent Libra) operating at 2 kHz repetition rate. The pump beam was generated by a non-collinear optical parametric amplifier (NOPA) which outputs tunable broadband pulses[25]. The pump was centred around 2.25 eV and compressed to ~ 20 fs using chirped mirrors. Details on pump fluence are reported in figure captions. The probe beam was obtained by white light continuum generation in a sapphire plate. Pump and probe beams were incident on the sample at near-normal incidence on the surface corresponding to the (001) crystallographic plane and cross polarized to avoid scattering artefacts. All spectra were dechirped taking into account the dispersion in the probe beam. The measured pump-probe data include both the coherent and incoherent dynamics, i.e., the total transient response of the sample. To isolate only the oscillatory (coherent) component for further analysis, a monoexponential fit of the incoherent dynamics was made to each wavelength (photon energy) with free parameters for the 512 data rows of each wavelength bin of our CCD. The monoexponential decays were then subtracted from the measured data to produce the map in Fig.2a. The FFT analysis in Fig.3 was performed on the exponential subtracted data in Fig.2a. Steady-state reflectivity was performed with a custom modified setup[26] based on an Agilent Cary 5000 spectrophotometer with single crystals held in vacuum at 77 K. A UV-enhanced aluminium mirror was used as a reference. Density Functional Perturbation Theory (DFPT)[27] with a plane wave basis was used to calculate the frequencies of the phonon modes at the Brillouin zone centre. We used the QUANTUM ESPRESSO package[28] with ultrasoft[29] or projector augmented-wave[30] pseudopotentials, and the local density approximation (LDA) was used with a Perdew-Zunger exchange-correlation functional[31]. Comparison of different exchange-correlation functional choices shows that this is likely to be an adequate level of approximation, especially for the lattice dynamics[32]. Full details of the DFT calculations are reported in the SI.

**Acknowledgments.**

Computational work was performed on the University of Bath's High Performance Computing Facility and was also supported by the University of Bath Cloud Pilot Project. Support for this work was provided by EPSRC grant EP/L015544. EC acknowledges funding from Italian PRIN project 2017BZPKSZ. We wish also to thank the Royal Society for support under the Wolfson Laboratory Refurbishment scheme. We appreciate the technical support by Phil Jones, Ash Moore and Clare Cambridge at University of Bath.

**References**

[1] a) T. Dekorsy, G. C. Cho, H. Kurz, in *Light Scattering in Solids Viii: Fullerenes, Semiconductor Surfaces, Coherent Phonons*, Vol. 76 (Eds: M. Cardona, G. Guntherodt) **2000**, p. 169; b) J. Shah, *Ultrafast spectroscopy of semiconductors and semiconductor nanostructures*, Springer, Heidelberg, Germany **1999**.

[2] Q. Wang, R. W. Schoenlein, L. A. Peteanu, R. A. Mathies, C. V. Shank, *Science* **1994**, 266, 422.

[3] J. Feldmann, K. Leo, J. Shah, D. A. B. Miller, J. E. Cunningham, T. Meier, G. Vonplessen, A. Schulze, P. Thomas, S. Schmittrink, *Phys. Rev. B* **1992**, 46, 7252.

[4] T. Dekorsy, H. Auer, C. Waschke, H. J. Bakker, H. G. Roskos, H. Kurz, V. Wagner, P. Grosse, *Phys. Rev. Lett.* **1995**, 74, 738.

[5] A. M. Weiner, D. E. Leaird, G. P. Wiederrecht, K. A. Nelson, *Science* **1990**, 247, 1317.

[6] a) C. Giannetti, M. Capone, D. Fausti, M. Fabrizio, F. Parmigiani, D. Mihailovic, *Adv. Phys.* **2016**, 65, 58; b) R. Mankowsky, B. Liu, S. Rajasekaran, H. Y. Liu, D. Mou, X. J. Zhou, R. Merlin, M. Forst, A. Cavalleri, *Phys. Rev. Lett.* **2017**, 118, 116402; c) S. Gerber, S.-L. Yang, D. Zhu, H. Soifer, J. A. Sobota, S. Rebec, J. J. Lee, T. Jia, B. Moritz, C. Jia, A. Gauthier, Y. Li, D. Leuenberger, Y. Zhang, L. Chaix, W. Li, H. Jang, J.-S. Lee, M. Yi, G. L. Dakovski, S. Song, J. M. Glownia, S. Nelson, K. W. Kim, Y.-D. Chuang, Z. Hussain, R. G. Moore, T. P. Devereaux, W.-S. Lee, P. S. Kirchmann, Z.-X. Shen, *Science* **2017**, 357, 71; d) D. Werdehausen, T. Takayama, M. Höppner, G. Albrecht, A. W. Rost, Y. Lu, D. Manske, H. Takagi, S. Kaiser, *Science Advances* **2018**, 4, eaap8652.

[7] a) T. Rohwer, S. Hellmann, M. Wiesenmayer, C. Sohrt, A. Stange, B. Slomski, A. Carr, Y. W. Liu, L. Miaja-Avila, M. Kallane, S. Mathias, L. Kipp, K. Rossnagel, M. Bauer, *Nature* **2011**, 471, 490; b) M. Porer, U. Leierseder, J. M. Menard, H. Dachraoui, L. Mouchliadis, I. E. Perakis, U. Heinzmann, J. Demsar, K. Rossnagel, R. Huber, *Nature Mater.* **2014**, 13, 857; c) S. Hellmann, T. Rohwer, M. Kallane, K. Hanff, C. Sohrt, A. Stange, A. Carr, M. M. Murnane, H. C. Kapteyn, L. Kipp, M. Bauer, K. Rossnagel, *Nature Comm.* **2012**, 3, 1069; d) C. J. Sayers, H. Hedayat, A. Ceraso, F. Museur, M. Cattelan, L. S. Hart, L. S. Farrar, S. Dal Conte, G. Cerullo, C. Dallera, E. Da Como, E. Carpene, *Phys. Rev. B* **2020**, 102, 161105; e) H. Hedayat, C. J. Sayers, D. Bugini, C. Dallera, D. Wolverson, T. Batten, S. Karbassi, S. Friedemann, G. Cerullo, J. van Wezel, S. R. Clark, E. Carpene, E. Da Como, *Phys. Rev. Res.* **2019**, 1, 11.

[8] a) K. Rossnagel, *J. Phys.-Condes. Matter* **2011**, 23, 213001; b) P. Monceau, *Adv. Phys.* **2012**, 61, 325.

[9] G. H. Han, D. L. Duong, D. H. Keum, S. J. Yun, Y. H. Lee, *Chem. Rev.* **2018**, 118, 6297.

[10] J. A. Wilson, F. J. Di Salvo, S. Mahajan, *Adv. Phys.* **1975**, 24, 117.

[11] a) P. Knowles, B. Yang, T. Muramatsu, O. Moulding, J. Buhot, C. J. Sayers, E. Da Como, S. Friedemann, *Phys. Rev. Lett.* **2020**, 124, 6; b) C. J. Sayers, L. S. Farrar, S. J. Bending, M. Cattelan, A. J. H. Jones, N. A. Fox, G. Kociok-Köhn, K. Koshmak, J. Laverock, L. Pasquali, E. Da Como, *Physical Review Materials* **2020**, 4, 025002.

[12] C. Sohrt, A. Stange, M. Bauer, K. Rossnagel, *Faraday Discussions* **2014**, 171, 243.

[13] a) L. Perfetti, A. Georges, S. Florens, S. Biermann, S. Mitrovic, H. Berger, Y. Tomm, H. Hochst, M. Grioni, *Phys. Rev. Lett.* **2003**, 90, 4; b) Y. Chen, W. Ruan, M. Wu, S. J. Tang, H. Y. Ryu, H. Z. Tsai, R. Lee, S. Kahn, F. Liou, C. H. Jia, O. R. Albertini, H. Y. Xiong, T. Jia, Z. Liu, J. A. Sobota, A. Y. Liu, J. E. Moore, Z. X. Shen, S. G. Louie, S. K. Mo, M. F. Crommie, *Nature Phys.* **2020**, 16, 218.

[14] A. S. Ngankeu, S. K. Mahatha, K. Guilloy, M. Bianchi, C. E. Sanders, K. Hanff, K. Rossnagel, J. A. Miwa, C. B. Nielsen, M. Bremholm, P. Hofmann, *Phys. Rev. B* **2017**, 96.

[15] X. Shi, W. You, Y. Zhang, Z. Tao, P. M. Oppeneer, X. Wu, R. Thomale, K. Rossnagel, M. Bauer, H. Kapteyn, M. Murnane, *Science Advances* **2019**, 5, eaav4449.

[16] S. Sugai, *Phys. Stat. Sol. B-Basic Research* **1985**, 129, 13.

[17] S. Sugai, K. Murase, S. Uchida, S. Tanaka, *Physica B & C* **1981**, 105, 405.

[18] H. J. Zeiger, J. Vidal, T. K. Cheng, E. P. Ippen, G. Dresselhaus, M. S. Dresselhaus, *Phys. Rev. B* **1992**, 45, 768.

[19] A. H. Reshak, S. Auluck, *Physica B* **2005**, 358, 158.

[20] A. R. Beal, H. P. Hughes, W. Y. Liang, *Journal of Physics C-Solid State Physics* **1975**, 8, 4236.

[21] A. Mann, E. Baldini, A. Odeh, A. Magrez, H. Berger, F. Carbone, *Phys. Rev. B* **2016**, 94, 115122.

[22] T. E. Stevens, J. Kuhl, R. Merlin, *Phys. Rev. B* **2002**, 65.

[23] X. B. Feng, J. Henke, C. Morice, C. J. Sayers, E. Da Como, J. van Wezel, E. van Heumen, *Phys. Rev. B* **2021**, 104.

[24] Y. C. Zhang, X. Shi, W. J. You, Z. S. Tao, Y. G. Zhong, F. C. Kabeer, P. Maldonado, P. M. Oppeneer, M. Bauer, K. Rossnagel, H. Kapteyn, M. Murnane, *Proc. Natl. Acad. Sci. USA* **2020**, 117, 8788.

[25] D. Brida, C. Manzoni, G. Cirmi, M. Marangoni, S. Bonora, P. Villoresi, S. De Silvestri, G. Cerullo, *J. Opt.* **2010** 12, 013001.

[26] J. W. R. Macdonald, G. Piana, M. Comin, E. von Hauff, G. Kociok-Kohn, C. Bowen, P. Lagoudakis, G. D'Avino, E. Da Como, *Mater. Horizons* **2020**, 7, 2951.

[27] P. Giannozzi, S. De Gironcoli, P. Pavone, S. Baroni, *Phys. Rev. B* **1991**, 43, 7231.

[28] P. Giannozzi, O. Andreussi, T. Brumme, O. Bunau, M. B. Nardelli, M. Calandra, R. Car, C. Cavazzoni, D. Ceresoli, M. Cococcioni, N. Colonna, I. Carnimeo, A. Dal Corso, S. de Gironcoli, P. Delugas, R. A. DiStasio, A. Ferretti, A. Floris, G. Fratesi, G. Fugallo, R. Gebauer, U. Gerstmann, F. Giustino, T. Gorni, J. Jia, M. Kawamura, H. Y. Ko, A. Kokalj, E. Kucukbenli, M. Lazzeri, M. Marsili, N. Marzari, F. Mauri, N. L. Nguyen, H. V. Nguyen, A. Otero-de-la-Roza, L. Paulatto, S. Ponce, D. Rocca, R. Sabatini, B. Santra, M. Schlipf, A. P. Seitsonen, A. Smogunov, I. Timrov, T. Thonhauser, P. Umari, N. Vast, X. Wu, S. Baroni, *J. Phys.-Condes. Matter* **2017**, 29.

[29] D. Vanderbilt, *Phys. Rev. B* **1990**, 41, 7892.

[30] a) P. E. Blochl, *Phys. Rev. B* **1994**, 50, 17953; b) G. Kresse, D. Joubert, *Phys. Rev. B* **1999**, 59, 1758.

[31] J. P. Perdew, A. Zunger, *Phys. Rev. B* **1981**, 23, 5048.

[32] L. S. Hart, S. M. Gunasekera, M. Mucha-Kruczynski, J. L. Webb, J. Avila, M. C. Asensio, D. Wolverson, *Phys. Rev. B* **2021**, 104.

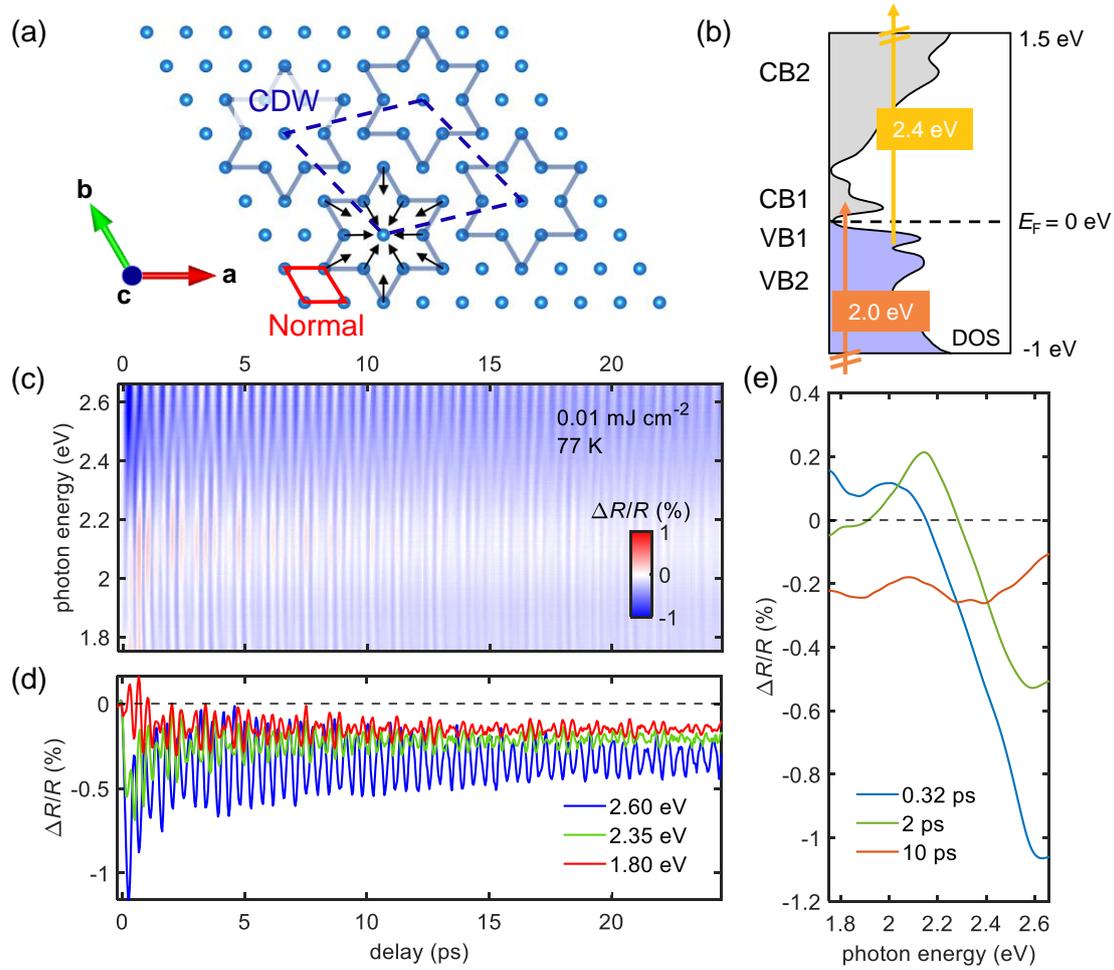

**Figure 1**. (a) Projection of the Tantalum atoms in a single layer of 1$T$-TaSe$_2$ along the $c$- axis, Selenium atoms have been removed for clarity. The hexagonal unit cell of the normal structure is indicated in red, while the √13 x √13 CDW superstructure unit cell is identified by the dashed blue line. 13 atoms star clusters are shown for the CDW unit cell only. (b) Scheme of energy bands showing a DOS adapted from ref. [13b] in proximity of the Fermi level, $E_F$, together with an illustration of the proposed final and initial bands involved in interband transitions for photons with energy in the range of our probe 1.75-2.65 eV (vertical arrows). (c) Transient reflectivity map of ΔR/R as a function of pump-probe time delay and probe photon energy. (d) Time traces extracted from panel (c) at specific probe photon energies. (e) ΔR/R spectra at specific time delays extracted from panel (c).

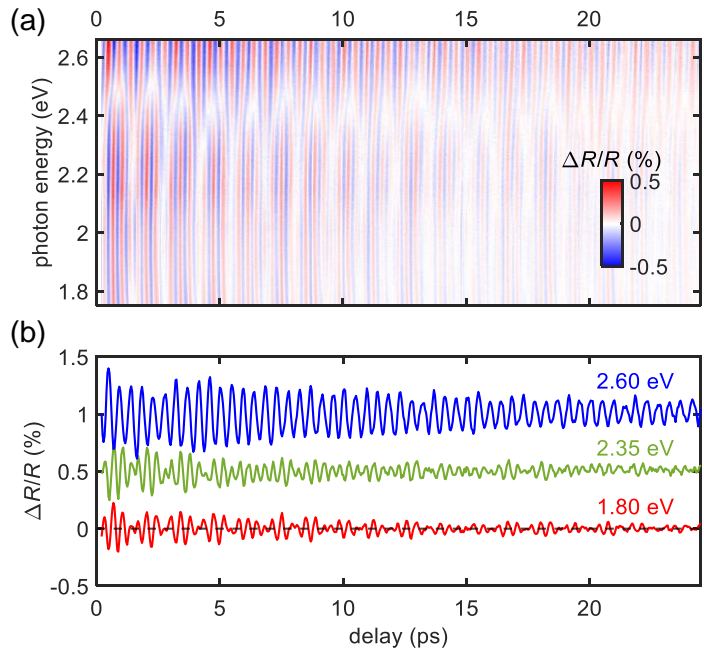

**Figure 2**. (a) Oscillatory component of the ΔR/R signal as a function of pump-probe time delay and probe photon energy after subtraction of the incoherent dynamics. (b) Time traces extracted from panel (a) at specific photon energies. The traces have been vertically offset for clarity.

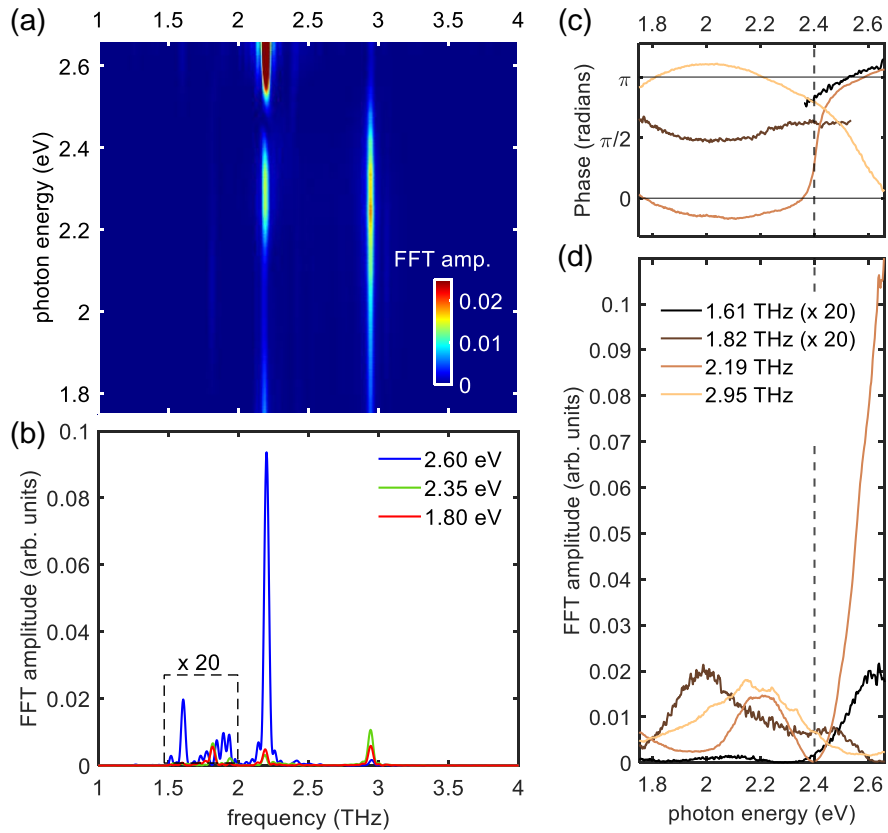

**Figure 3**. (a) Fast Fourier Transform (FFT) amplitude map. (b) FFT amplitude as a function of frequency for specific photon energies, data between 1.5 and 2.0 THz have been multiplied by a factor of 20 for clarity. (c) Phase of coherent phonon modes as a function of probe photon energy. (d) Spectral dependence of the FFT amplitude of the four most prominent modes. The amplitude of the 1.61 THz and 1.82 THz modes has been multiplied by a factor of 20. The vertical dashed line corresponds to 2.4 eV.

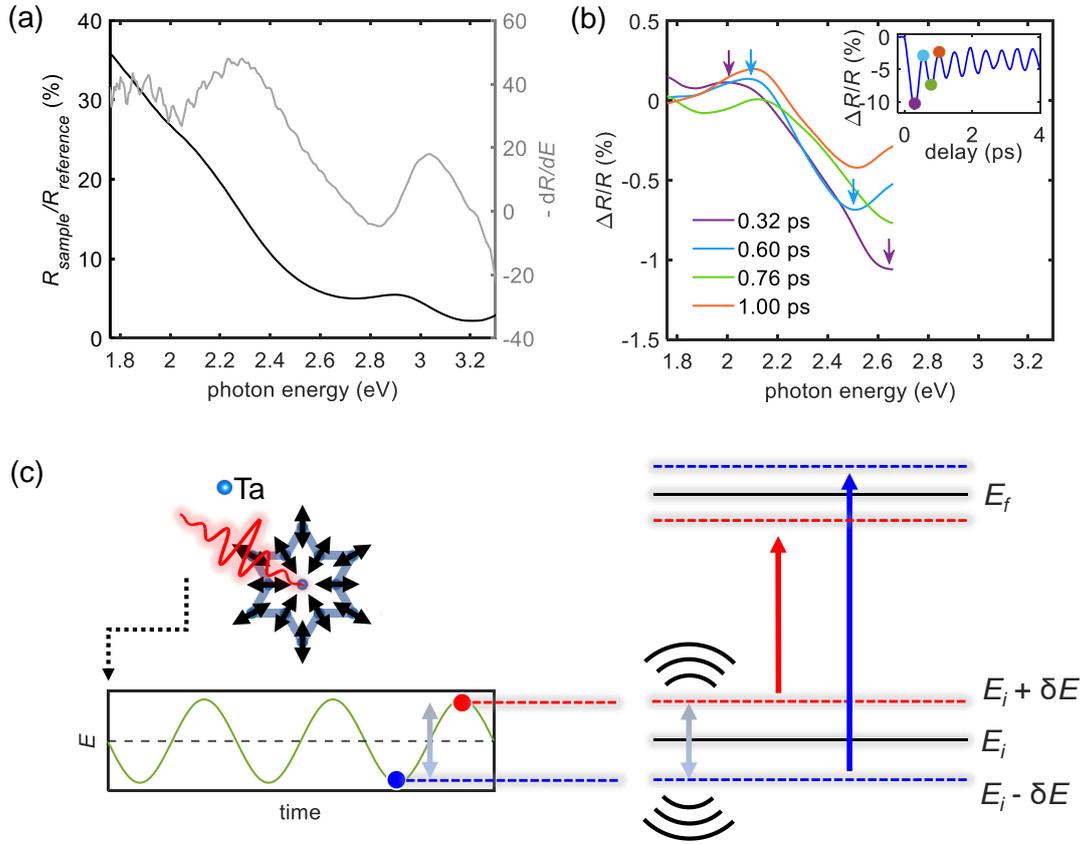

**Figure 4**. (a) Steady-state reflectivity spectrum of 1$T$-TaSe$_2$ at 77 K (black curve, left axis) The grey curve (right axis) is the negative first derivative of the steady-state reflectivity. (b) ΔR/R spectra at selected pump-probe time delays indicated in the legend corresponding to the maxima and minima in the coherent oscillations of the time trace at 2.6 eV shown in the inset. The arrows indicate maxima and minima of the spectral features. (c) Oscillations in the binding energy of initial and final states in an interband transition. The right panel shows how an optical transition involving such bands is affected and the corresponding out of phase oscillations in the optical response at photon energies below (red arrow) and above (blue arrow) the centre energy.